\documentclass[twocolumn]{aastex62}
\usepackage{graphics,epsf}
\usepackage{amsmath}                
\usepackage{amsfonts}               
\usepackage{amssymb}                
\usepackage{epsfig}                 
\usepackage{appendix}
\usepackage{graphicx}
\usepackage{float}
\usepackage{color}
\usepackage{multirow}
\usepackage[para,online,flushleft]{threeparttable}

\newcommand{\yr}{{~\rm yr}}

\newcommand{\AU}{{~\rm AU}}

\begin{document}

\title{On the role of reduced wind mass-loss rate in enabling exoplanets to shape planetary nebulae}

\email{ahlam.hegazi@campus.technion.ac.il; ealealbh@gmail.com; \\ soker@physics.technion.ac.il}

\author{Ahlam Hegazi}
\affiliation{Department of Physics, Technion – Israel Institute of Technology, Haifa 3200003, Israel}

\author{Ealeal Bear}
\affiliation{Department of Physics, Technion – Israel Institute of Technology, Haifa 3200003, Israel}

\author{Noam Soker}
\affiliation{Department of Physics, Technion – Israel Institute of Technology, Haifa 3200003, Israel}
\affiliation{Guangdong Technion Israel Institute of Technology, Guangdong Province, Shantou 515069, China}

\begin{abstract}
We use the stellar evolution code \textsc{MESA-binary} and follow the evolution of {{{{ three exoplanets and two brown dwarfs (BDs) }}}} to determine their potential role in the future evolution of their parent star on the red giant branch (RGB) and on the asymptotic giant branch (AGB). We limit this study to exoplanets {{{{ and BDs }}}} with orbits that have semi-major axis of $1 \AU \la a_0 \la 20 \AU$, a high eccentricity, $e_0 \ga 0.25$, and having a parent star of mass $M_{\ast, 0} \ge 1 M_\odot$. We find that the star HIP~75458 will engulf its planet HIP~75458~b during its RGB phase. The planet will remove the envelope and terminate the RGB evolution, leaving a bare helium core of mass $0.4 M_\odot$ that will evolve to form a helium white dwarf. Only in one system out of five, the planet beta~Pic~c will enter the envelope of its parent star during the AGB phase. For that to occur, we have to reduce the wind mass-loss rate by a factor of about four from its commonly used value. 
This strengthens an early conclusion, which was based on exoplanets with circular orbits, that states that to have a non-negligible fraction of AGB stars that engulf planets we should consider lower wind mass-loss rates of isolated AGB stars (before they are spun-up by a companion). Such an engulfed planet might lead to the shaping of the AGB mass-loss geometry to form an elliptical planetary nebula. 
\end{abstract}


\section{Introduction} 
\label{sec:intro}

Hundreds of observational and theoretical studies in recent years have converged on the understanding that binary interaction shapes the majority, and possibly all, planetary nebulae (PNe) 
(e.g., limiting to a sample from the last five years, \citealt{Jonesetal2016, Chiotellisetal2016, Akrasetal2016, GarciaRojasetal2016, Jones2016, Hillwigetal2016a, Bondetal2016, Chenetal2016, Madappattetal2016, Alietal2016, Hillwigetal2016b,  JonesBoffin2017b, Barker2018, BondCiardullo2018, Bujarrabaletal2018, Chenetal2018, Danehkaretal2018, Franketal2018, GarciaSeguraetal2018, Hillwig2018,  MacLeodetal2018, Miszalskietal2018PASA, Sahai2018ASPC, Wessonetal2018, Alleretal2019, Desmursetal2019, Jones2019H, Kimetal2019, Kovarietal2019, Miszalskietal2019MNRAS487, Oroszetal2019, Akrasetal2020, BermudezBustamanteetal2020, Jones2020CEE}). 
Substantially smaller number of studies deal with the possibility that planets and brown dwarfs (BDs) might also shape PNe (e.g, \citealt{DeMarcoSoker2011, Kervellaetal2016, Boyle2018PhDT, SabachSoker2018a, SabachSoker2018b, Schaffenrothetal2019}, as some examples from the last decade).

A stellar companion can strongly deform the envelope of the asymptotic giant branch (AGB) progenitor of the PN, by spinning-up the envelope and by the direct effects of its gravity, mainly during a common envelope evolution (CEE) phase and during the termination of the CEE. One of the extreme outcomes at the termination of the CEE might be two opposite `funnels' along the symmetry axis of the bloated AGB envelope, which can collimate bipolar outflows
(e.g., \citealt{Soker1992, Reichardtetal2019, GarciaSeguraetal2020, Zouetal2020}). 
A stellar companion can also deform the envelope by accreting mass and launching jets during the CEE (e.g., \citealt{Chamandyetal2018a, LopezCamaraetal2019, Schreieretal2019, Shiberetal2019, LopezCamaraetal2020}). All these processes shape the descendent nebula to possess large-scale  highly non-spherical morphologies. Planet companions, on the other hand, are not expected to have these large effects. It is not clear yet whether planets can launch jets. Even if they do (e.g., \citealt{SokerWorkPlans2020}), the outcome might be two opposite small clumps along the symmetry axis (\textit{ansae}; FLIERS). 
  
It seems that the main effect of a planet in shaping the mass-loss toward a non-spherical PN is by spinning-up the envelope, to the degree that the mass-loss becomes axisymmetrical. \cite{Soker1998AGB} discussed the way by which a planet can enhance mass-loss and can lead to a non-spherical outflow from giant stars, like AGB stars, or red giant branch (RGB) stars. It goes as follows. A planet-spun-up AGB envelope might sustain a dynamo  (e.g., \citealt{NordhausBlackman2006}), that in turn leads to non-spherical mass-loss by the effect of magnetic fields (e.g., \citealt{LealFerreiraetal2013, Vlemmings2018}), including possibly the influence on dust formation and distribution (e.g., \citealt{Soker2000, Soker2001a, Khourietal2020}). Another effect of massive planets that are deep inside the envelope of giant stars, after the dynamical in-spiral phase, is the excitation of waves that become large on the surface and might influence the rate and morphology of dust formation and therefore of the outflow (e.g., \citealt{Soker1993}). On a more general ground, dust formation seems to be an important process in the last phases of the CEE, both for sub-stellar and stellar companions 
(e.g., \citealt{Soker1992b, Soker1998AGB, GlanzPerets2018, Iaconietal2019, Iaconi2020}).

Stars on the upper RGB and AGB can acquire a large amount of angular momentum by engulfing planets
(e.g., \citealt{Soker1996, SiessLivio1999AGB, Massarotti2008, Carlbergetal2009, VillaverLivio2009, MustillVillaver2012, Nordhausetal2010, NordhausSpiegel2013, GarciaSeguraetal2014, Staffetal2016, AguileraGomezetal2016, Veras2016, SabachSoker2018a, SabachSoker2018b}).
The probability for this process to take place on the upper AGB sensitively depends on the mass-loss rate from the star. 
{{{{ We note that most known exoplanets will be engulfed before  their parent star reaches the upper RGB, because most known exoplanets have close orbits (detection biased). These systems are not relevant to us as the planet will not shape the outflow just before the termination  the RGB or AGB. Exoplanets (and BDs) that shape post-RGB nebulae or PNe should have semi-major axis of $a_0 \ga 0.5 \AU$.  }}}}

In earlier studies our group considered the possibility that AGB stars that did not (yet) interact with any companion that substantially spun-up the envelope, have much lower wind mass-loss rates than what traditional formulae give \citep{SabachSoker2018a, SabachSoker2018b}. We termed these angular momentum $(\vec{J}~)$ isolated stars {\it Jsolated stars}.

\cite{SabachSoker2018a} study the fate of four observed exoplanetary systems that have low eccentricity. To follow the evolution of the star they use the single mode of the evolutionary code \textsc{MESA} (section \ref{sec:method}). To examine whether tidal forces will cause the planet to spiral-in to the envelope of the star during the AGB phase, they use a simple prescription for the tidal force. \cite{SabachSoker2018a} found that when low-mass stars evolve with the traditional wind mass-loss rate they are not likely to swallow their planets in these four systems. 
With a lower mass-loss rate, down to about $15 \%$ of the traditional one, the stars reach much larger radii on their AGB and much larger luminosities. The larger radii substantially increase the likelihood for the AGB star to swallow the planet. This, by the studies we cited above, might lead to the formation of elliptical PNe. The higher luminosity might account for bright PNe in old stellar population (see relevant discussion in \citealt{SabachSoker2018a, SabachSoker2018b}. 

\cite{SabachSoker2018a} justified the much lower wind mass-loss rate by their assumption that the stellar samples from which the mass-loss rate formulae were derived were contaminated by AGB stars that suffer binary interaction, and binary interaction increases the mass-loss rate.  {{{{ About half of main sequence stars in the mass range $1-2 M_\odot$ have a stellar binary companion (e.g., \citealt{MoeDiStefano2017}). Many of these binary systems are close enough for the companion to increases the mass-loss rate of the primary star. The point that \cite{SabachSoker2018a} make is that for many other stars a close exoplanet (or a close BD) enhances the mass-loss rate. Overall, both stellar and sub-stellar companions enhance the mass-loss rate from many AGB stars. }}}}

Specifically for low mass companions, down to planets, \cite{SabachSoker2018b} suggest that giant stars that acquire no angular momentum from a companion along their late evolution (beyond the main sequence), i.e., Jsolated stars, have much lower mass-loss rates than what traditional formulae give. 
AGB stars with lower mass-loss rates reach much higher luminosities in the post-AGB track, when they ionise the PN. 
\cite{SabachSoker2018b} further argue that it might be that the bright PNe in old stellar populations result from the combination of lower mass-loss rates that they explored, and of higher AGB luminosities that some new stellar models give (e.g., \citealt{Bertolami2016, Gesickietal2018, Mendez2017, Reindletal2017}). 

We here adopt the approach of \cite{SabachSoker2018a} and \cite{SabachSoker2018b} in considering a much lower wind mass-loss rate (section \ref{sec:method}). We differ from them by studying the fate of observed exoplanets {{{{ and BDs }}}} with high eccentricities, for which we must use the binary mode of \textsc{MESA} to follow the evolution of the planetary systems (section \ref{sec:method}). 
Our study is another in a series of papers that study the fate of confirmed exoplanets as their parent stars turn to RGB or AGB stars (e.g., \citealt{NordhausSpiegel2013, SabachSoker2018a}). We describe the results of our simulations in section \ref{sec:Evolution}, and we summarise our main results in section \ref{sec:Summary}. 
   
\section{NUMERICAL METHOD}
\label{sec:method}

\subsection{The \textsc{MESA-binary} code}
\label{subsec:MesaBinary}

To follow the fate of the five observed systems, {{{{ three exoplanets and two BDs, }}}} that we study here, we conduct stellar evolution simulations using the Modules for Experiments in Stellar Astrophysics (MESA; \citealt{Paxtonetal2011, Paxtonetal2013, Paxtonetal2015, Paxtonetal2018, Paxtonetal2019}), version 10398 in its binary mode. 
In each system we follow the evolution of the parent star, either to the time the star engulfs its planet {{{{ (or BD; in what follows in many cases we refer by planets also to BDs) }}}} and the system enters a CEE phase, or to its post-AGB phase if no engulfment takes place.  

We study planets with high-eccentricity orbits and so we have to pay attention to tidal forces that act to circularise and synchronise the orbit (the later effect results in a decrease in the semi-major axis). We set the tidal effects in \textsc{MESA-binary} (the parameters \textit{do\_tidal\_circ} and \textit{do\_tidal\_sync}), taking the circularization type \textit{'Hut\_conv'} which is the default of \textsc{MESA-binary} for convective envelope \citep{Hurleyetal2002}. This is relevant to our study as the planets we follow experience strong tidal interaction only during the RGB and AGB phases of their parent stars, when the envelope is fully convective. 
We turn off the effects of magnetic breaking (the parameter \textit{do\_jdot\_mb}) as we expect weak magnetic activity during the RGB and AGB phases before the planet enters the envelope. We take all other parameters to have their default values in \textsc{MESA-binary}. 

\subsection{Mass-loss scheme}
\label{subsec:MassLoss}

As we mentioned above, we adopt our earlier approach (\citealt{SabachSoker2018a, SabachSoker2018b}, where there are more details and discussions of the low wind mass-loss rate), and give here only the essential information.
For the empirical mass-loss formula for winds on the RGB we take \citep{Reimers1975} 
\begin{equation}
\dot{M_{\rm RGB}}= 4\times10^{-13} \eta  LM^{-1} R,
\label{eq:Reimers}
\end{equation}
where the giant's mass, $M$, luminosity $L$, and radius $R$, are in solar units, and $\eta$ is the wind mass-loss rate efficiency parameter. 
{{{{ The mass-loss rate on the upper AGB should be larger than the Reimer formula (e.g., \citealt{VassiliadisWood1993}). Therefore, we use the MESA formula from \cite{Bloecker1995}
\begin{equation}
\dot{M_{\rm AGB}}= 4.83 \times10^{-9} M^{-2.1} L^{2.7} {\dot M_{\rm RGB}} .
\label{eq:Bloecker}
\end{equation}
 }}}}
We follow \citealt{SabachSoker2018a} and take the same value of $\eta$ for the mass-loss rate expressions on the RGB and on the AGB. 

The commonly used value is $\eta=0.5$ (e.g., \citealt{McDonaldZijlstra2015}). With the assumption that Jsolated stars (those that acquired no angular momentum from a companion) experience a much lower wind mass-loss rate than non-Jsolated stars, we also take lower values of $\eta$. 
\cite{Miglioetal2012}, for example, find for the old metal-rich cluster NGC 6791, that this parameter might be as low as $\eta=0.1$, i.e., much lower than typically taken. We follow \cite{SabachSoker2018a} and study the range $0.05 \le \eta \le 0.5$.
 
One observational finding is directly relevant to our study that aims at the shaping of elliptical PNe. This finding is the observation that many elliptical PNe have an outer faint and spherical halos (e.g.  \citealt{Corradietal2003}). Since single AGB (Jsolated) stars spin extremely slowly on the upper AGB (e.g., \citealt{Soker2006b}), we expect these stars to blow a spherically faint halo. Interacting with a low mass companion on the upper AGB causes these stars to have a non-spherical mass-loss and at a higher mass-loss rate, forming the brighter elliptical inner shell  (e.g., \citealt{Soker2000}). These PNe might suggest a late interaction with a very low mass companion, e.g., a BD or a planet.
 
\section{Evolution of five observed exoplanets and drown dwarfs}
\label{sec:Evolution}
\subsection{The five systems}
\label{subsec:FiveSystems}

Our aim is to explore which of the five observed exoplanets and BDs that we list in table \ref{tab:FiveExoplanets} might enter the envelope of their parent star when the later is on its upper AGB, and for what wind mass-loss rate efficiency parameter $\eta$. We study these specific systems that we found by searching the Extrasolar Planets Encyclopaedia; (exoplanet.eu; \citealt{Schneideretal2011}; {{{{ the system HD~72946 with a BD companion is from \citealt{Maireetal2020}) }}}} because they have the relevant range of all parameters, in particular a semi-major axis in the range of $1 \AU \la a \la 20 \AU$. There are many more exoplanets with a semi-major axis in this range, but the mass of the planet and/or the eccentricity are not known. The first five columns of the table list the name and input parameters from observations. We add a subscript `0' to indicate the initial values of the stellar mass $M_{\ast}$, of the semi-major axis $a$, and of the eccentricity $e$, as these quantities change during the post-main sequence evolution. We do not change the planet mass $M_{\rm p}$ {{{{ (by planet we refer below also to the two WDs) }}}} during the evolution. {{{{ The giant star will not evaporate the planet (e.g., \citealt{Schreiberetal2019}), and the planet will accrete almost no mass before it enters the envelope. }}}} The last six columns of table \ref{tab:FiveExoplanets} indicate the outcome for six different values of the wind mass-loss rate parameter $\eta$ (equations \ref{eq:Reimers} and \ref{eq:Bloecker}). We either indicate that the star does not engulf the planet, and so there is no CEE (`No CEE'), or in cases where the planet does enter a CEE, we indicate the core mass, $M_{\rm core}$, and the envelope mass, $M_{\rm env}$ at the onset of the CEE. 
\begin{table*} 
\centering
\begin{tabular}{|l|c|c|c|c|c|c|c|c|c|c|}
\hline
Planet/BD & $M_{\ast,0}$ & $M_{\rm p}$ & $a_0$ & $e_0$ & \multicolumn{6}{c|}{Outcome as function of $\eta$: No CEE or $M_{\rm core};M_{\rm env}$} \\ \hline
 & $M_\odot$ & $M_{\rm J}$ & $R_\odot$ (AU) &  & $\eta=0.5$ & $\eta=0.15$ & $\eta=0.12$ & $\eta=0.09$ & $\eta=0.07$ & $\eta=0.05$ \\ \hline
beta~Pic~c & 1.73 & 9.37 & 585 (2.72)& 0.248 & No CEE & No CEE & 0.583;0.595 & 0.581;0.87 & 0.579;0.96 &  \\ \hline
HR 5183 b & 1.07 & 3.2 & 3870 (18)& 0.84 & No CEE &  &  &  & No CEE & No CEE \\ \hline
HD 72946 & 1.01 & 72.5 & 1393 (6.45)& 0.49 & No CEE &  &  &  & No CEE & No CEE \\ \hline
HD 38529 c & 1.48 & 23.7 & 793 (3.695)& 0.36 & No CEE &  &  &  & No CEE & No CEE \\ \hline
HIP~75458~b & 1.4 & 9.4 & 273 (1.275)& 0.71 & 0.407;0.91 &  &  &  & 0.404;0.98 & 0.404;0.99 \\ \hline
\end{tabular}
\caption{The relevant properties and outcomes of the five systems {{{{ (including HD~72946 and HD~38529~c that are BDs). }}}} The first five columns list the input parameters, the planet name, the present primary star mass $M_{\ast , 0}$, the planet mass $M_{\rm p}$ in units of Jupiter mass $M_{\rm J}$, the present semi-major axis and the eccentricity of its orbit, $a_0$ and $e_0$, respectively. The right six columns list the outcome as function of six values of the mass-loss parameter, from $\eta=0.5$ (the common value) to $\eta=0.05$. We either indicate that the star does not engulf the planet and the system avoids a common envelope evolution (CEE), or in cases where the planet does get into a CEE we list the core mass and the envelope mass at the beginning of the CEE (in $M_\odot$). If for maximum ($\eta=0.5$) and minimum ($\eta=0.05$ or $\eta=0.07$) mass-loss rate parameters no engulfment occurs, there is no need to simulate the middle values, hence the empty boxes. }
\label{tab:FiveExoplanets}
\end{table*}
   
\subsection{Engulfment on the RGB}
\label{subsec:RGB}

\begin{figure}
\includegraphics[trim= 5.2cm 5.5cm 3.5cm 7.1cm,clip=true,width=0.54\textwidth]{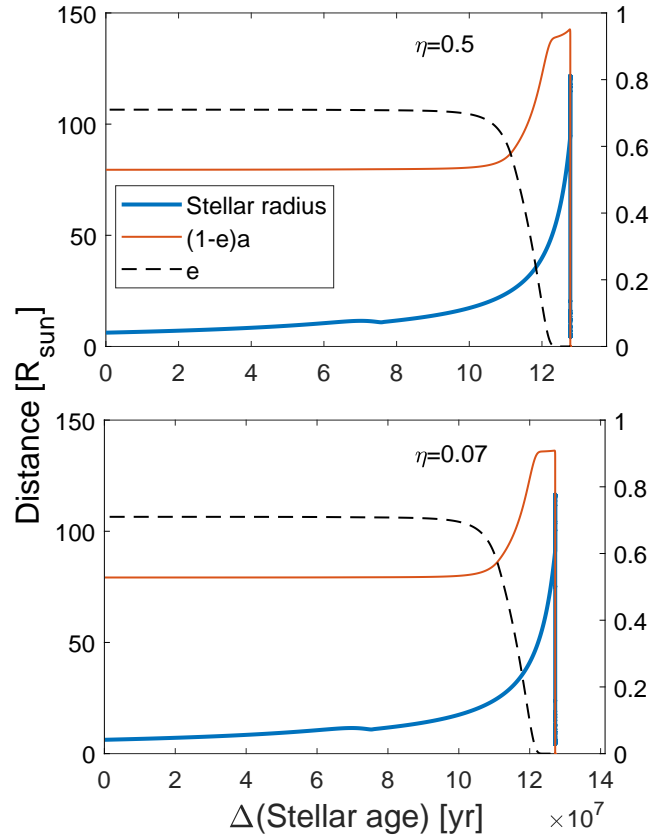}
\caption{The RGB evolution of the stellar radius (blue-thick line), the periastron distance (red line), and eccentricity (black-dashed line; scale on the right axis with $e_0=0.71$.) for the system HIP~75458 and for two values of the wind mass-loss rate efficiency parameter $\eta$. The upper and lower panels start at times (from the zero age main sequence) of $t=3.45\times 10^9 \yr$ and $t=3.435\times 10^9 \yr$, respectively. 
 }
\label{fig:HIP_75458b_TEST3}
\end{figure}

\begin{figure}
\includegraphics[trim= 5.2cm 5.5cm 3.5cm 6.9cm,clip=true,width=0.54\textwidth]{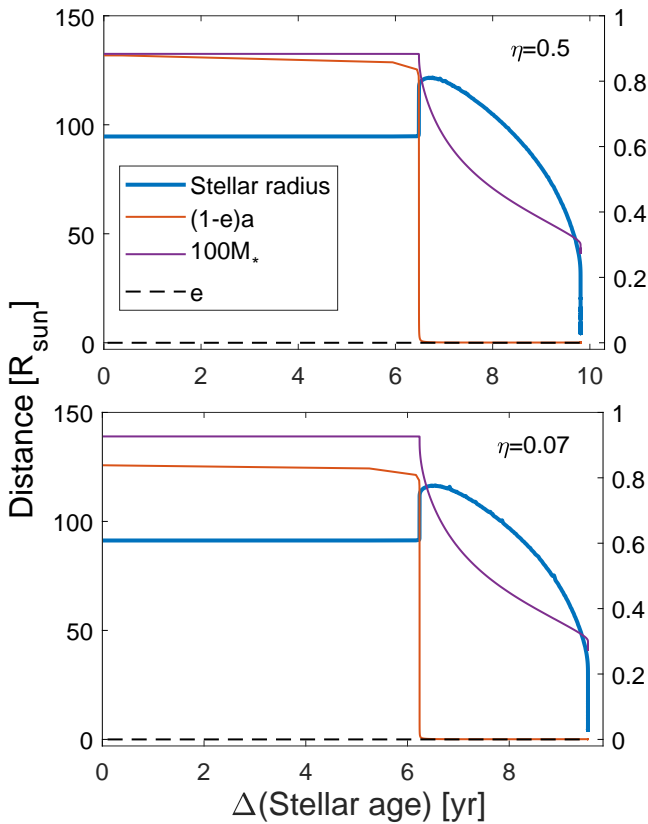}
\caption{Like Fig. \ref{fig:HIP_75458b_TEST3} but zooming on the formation of the CEE in the system HIP~75458 on the upper RGB. The upper and lowers panels start at $t=3.58 \times 10^9 \yr$ and at $t=3.56 \times 10^9 \yr$, respectively. The purple line is the stellar mass in units of $0.01M_\odot$, starting with $M_{\ast, 0}=1.4 M_\odot$ (corresponding to 140 on the vertical axis). Note that the orbit is circularised before the onset of the CEE.  }
\label{fig:HIP_75458b_Zoom}
\end{figure}

The planet HIP~75458~b enters the envelope of its parent star when the later is on the RGB for all values of $\eta$ that we use here. In Fig. \ref{fig:HIP_75458b_TEST3} we present the evolution of the stellar radius, periastron distance, and eccentricity of this system in the relevant post-main sequence phases. We see that tidal forces circularise the orbit before the onset of the CEE. Although the periastron distance increases, the semi-major axis decreases from its initial value of $a_0=273 R_\odot$ to about $a \simeq 140 R_\odot$, before it rapidly decreases as the planet dives into the RGB envelope. In Fig. \ref{fig:HIP_75458b_Zoom} we zoom on a time period of about $10 \yr$ when the planet enters the envelope of its parent star. We also present the evolution of stellar mass (purple line).  

The planet HIP~75458~b removes the envelope of its parent star during the RGB phase and leaves a bare helium core of mass $M_{\rm core}=0.4 M_\odot$, which then cools as a helium white dwarf. The planet might cause the nebula of the RGB star to have an elliptical shape. By definition this is not a PN. However, it is still a relevant system to our study.  The influence of planets on the evolution of RGB stars and on their later evolution to the horizontal branch has been the subject of a number of theoretical and observational papers  (e.g., \citealt{Soker1998HB, SiessLivio1999RGB, Carlbergetal2009, Geieretal2009, Heber2009ARAA, Charpinetetal2011, BearSoker2012, Silvottietal2014, Carlbergetal2016, Jimenezetal2020}). {{{{ Planets down to a mass of $M_{\rm p} \simeq M_{\rm J}$ might remove the entire hydrogen-rich envelope of their parent RGB star if they enter the envelope on the upper RGB. Lower mass planets are likely to be evaporated before they reach close to the core \citep{Soker1998HB}; they release less gravitational energy and therefore cannot unbind the entire envelope. 
}}}}

We find here that the system HIP~75458 belongs to a class of systems where the planet terminates the evolution of the star on the RGB, or at least causes the star to lose most of its envelope and to become a blue horizontal branch star \citep{Soker1998HB}.

\subsection{Possible shaping of PNe}
\label{subsec:PNe}
   
Not including HIP~75458~b that suffers RGB engulfment, we find that out of the other five exoplanets, only beta~Pic~c might enters a CEE during the AGB phase of its parent star (table \ref{tab:FiveExoplanets}). {{{{ First we note that the other planet in beta~Pic, the planet beta~Pic~b, has a semi-major axis of $9.68 \AU$ and a mass of $\simeq 12.7 M_{\rm J}$. With a semi-major axis that is about 3.5 larger than that of beta~Pic~c, and being only slightly more massive, we ignore the influence of beta~Pic~b on the evolution of beta~Pic~c that we study here. On the other hand, the close planet beta~Pic~b will induce non-spherical mass-loss geometry from the parent star if the planet enters the stellar envelope. Such a non-spherical mass-loss process will influence the orbit of the wider planet beta~Pic~b, as non-spherical mass-loss might do (e.g., \citealt{Verasetal2013, DosopoulouKalogera2016}). }}}}
  
We have to reduce the mass-loss rate by about a factor of four below the commonly used value $(\eta=0.5$) for beta~Pic~c to enter a CEE. 
\cite{SabachSoker2018a} find that in most cases they require $0.05 \la \eta \la 0.15$ for planets to enter a CEE with their parent star when the later is on its AGB. Our result for beta~Pic~c is compatible with their finding. 
 
\begin{figure}
\includegraphics[trim= 5.5cm 5.5cm 3.5cm 6.2cm,clip=true,width=0.6\textwidth]{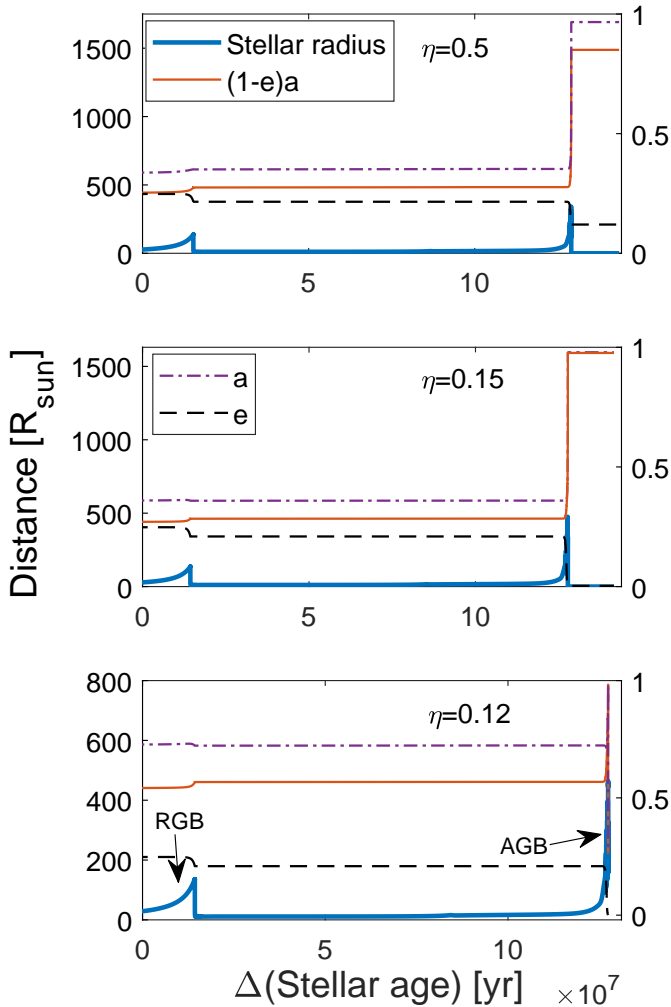}
\caption{The evolution of the stellar radius (blue-thick line), semi-major axis (dash-dotted-purple line; $a_0=585 R_\odot$), the periastron distance (red line), and eccentricity (black-dashed line; scale on the right axis with $e_0=0.24$) for the system beta~Pic~c and for three values of the wind mass-loss rate efficiency parameter $\eta$. The graphs include the RGB (first peak in radius), horizontal branch, and AGB (second peak in radius) phases of the evolution, and in the upper two panels the early post-AGB phase as well. The upper panel starts at $t=1.72 \times 10^9 \yr$, and the two lower panels at $t=1.71 \times 10^9 \yr$. Note the different scales of the three panels.}
\label{fig:beta_Picc_TEST3}
\end{figure}

We present the evolution of beta~Pic~c for three values of $\eta$ in Fig.  \ref{fig:beta_Picc_TEST3}. We notice that already on the RGB tidal interaction reduces somewhat the eccentricity. Then, during the AGB phase of the parent star when mass-loss rate is high, there are the competing effects of mass-loss that acts to increase the semi-major axis, and of tidal interaction that acts to circularise the orbit and to reduce orbital separation (as the spin of the AGB is much slower than the orbital motion of the planet). {{{{ Before the planet enters the envelope the mass-loss geometry is spherical on average. Therefore, the way the code \textsc{MESA-binary} treats the effect of mass-loss on the semi-major axis is accurate for our case.  }}}} For $\eta=0.5$ and $\eta = 0.15$ mass-loss rate is high, and the effect of mass-loss in enlarging the orbital separation wins that of the tidal interaction. In the case of $\eta=0.15$ the tidal force is strong enough to circularise the orbit. For $\eta=0.12$ the AGB reaches a larger radius on the AGB and, because tidal interaction is very sensitive to the ratio of the stellar radius to semi-major axis, tidal interaction manages to bring the planet into the AGB envelope.  

\begin{figure}
\includegraphics[trim= 5.0cm 5.5cm 2.5cm 6.5cm,clip=true,width=0.6\textwidth]{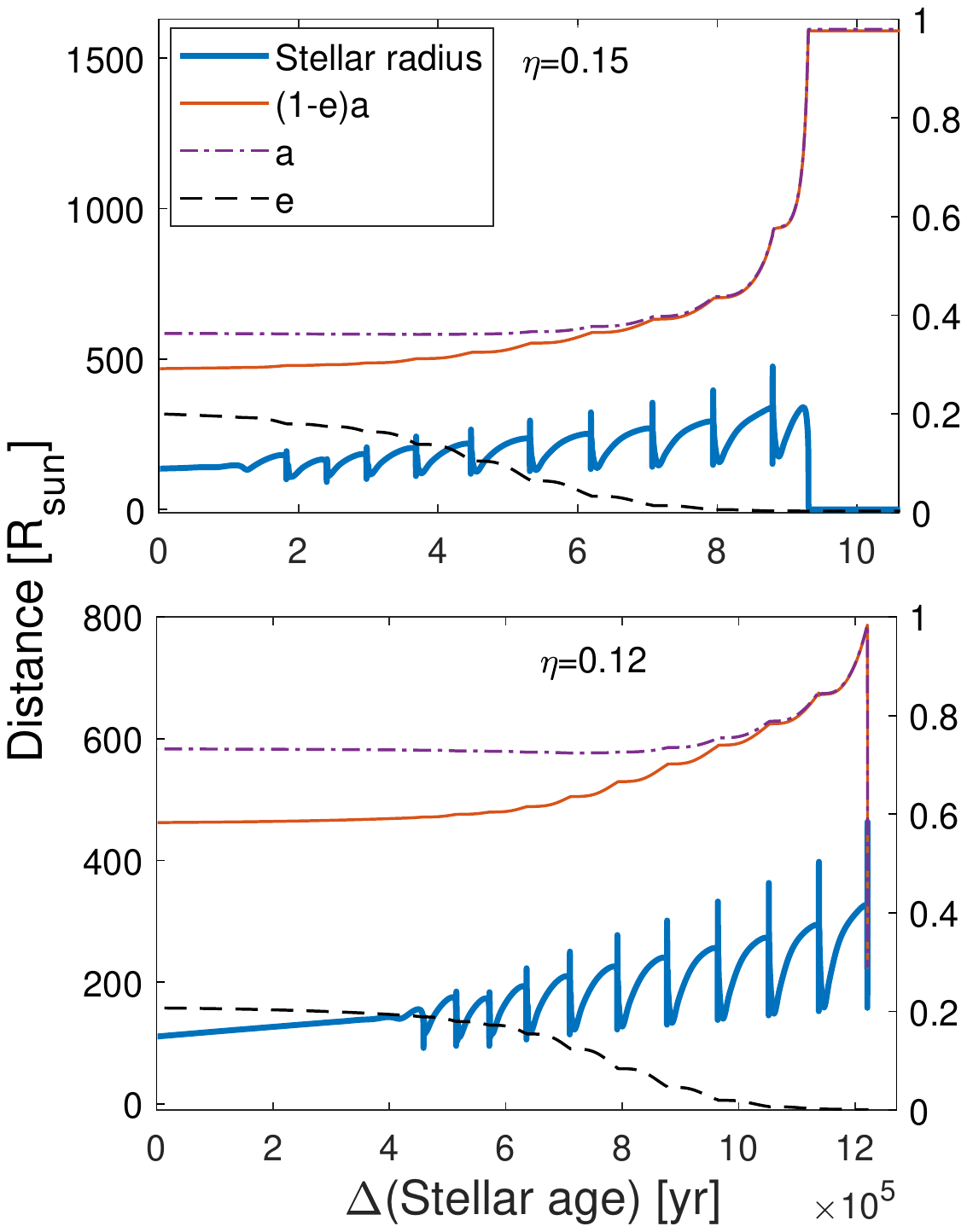}
\caption{Zooming on the final AGB evolution of the two lower panels of Fig. \ref{fig:beta_Picc_TEST3}. {{{{ The spikes in the blue line (stellar radius) are helium-shell flashes. }}}} The upper and lower panels start at $t= 1.84 \times 10^9 \yr$. Note the different scales of the two panels. }
\label{fig:ZoomedToAGB}
\end{figure}

In Figure \ref{fig:ZoomedToAGB} we zoom on the final million years or so of the evolution of the two lower panels of Fig. \ref{fig:beta_Picc_TEST3}. We see the helium-shell flashes effect in causing substantial envelope expansion. This increases the tidal interaction strength, that in turn slows down the increase in the semi-major axis, or even decreases it a little. The star finally engulfs the planet (lower panel) during such an envelope expansion of a helium-shell flash.   
   
Consider the possible role of the planet in shaping the descendant PN. The planet beta~Pic~c, of mass $M_{\rm p} = 9.3 M_{\rm J}$, enters the envelope when its mass is $M_{\rm env} \simeq 0.6-1 M_\odot$ for the values of $\eta$ that we use. Namely, the planet mass is 
$M_{\rm p} \simeq 0.01 M_{\rm env}$. Such a planet might excite large-amplitude (tens of per cents) oscillatory modes on the surface of the AGB star when it is deep inside the envelope (equation 5.7 in \citealt{Soker1992}), and might substantially spin-up the envelope (equation 10 in \citealt{Soker2001SpinUp}). 
   
We consider the planetary system of beta~Pic to be a future progenitor of an elliptical PN due to the expected entrance of the planet beta~Pic~c to a CEE during the AGB phase of its parent star.
      
\subsection{Examining the role of eccentricity}
\label{subsec:eccentricity}
  
\begin{figure}
\includegraphics[trim= 5.5cm 5.5cm 3.5cm 6.2cm,clip=true,width=0.6\textwidth]{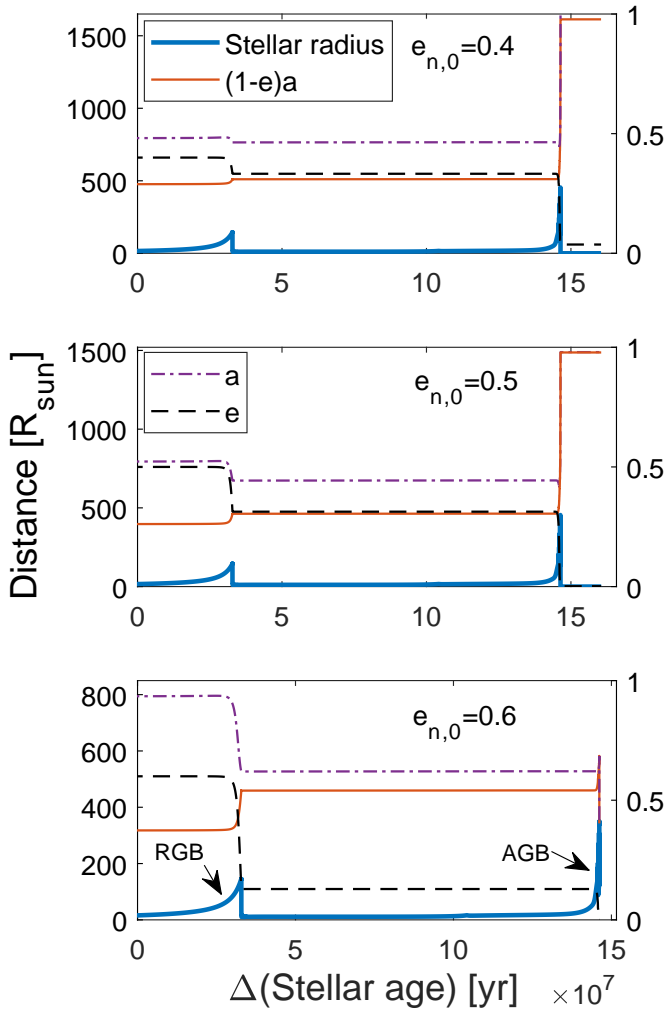}
\caption{The evolution with time of the stellar radius (thick-blue line), semi-major axis (dash-dotted-purple line; $a_0=793 R_\odot$), periastron distance (red line), and eccentricity (dashed-black line; scale on the right side) of the planet HD~38529~c, but with trial eccentricities, $e_{\rm n,0}$, that are larger than the observed value of $e_0=0.36$. All panels start at $t=2.92 \times 10^9 \yr$. The graphs include the RGB, horizontal branch, and AGB phases of the evolution, and in the upper two panels the early post-AGB phase as well. For all the simulations the efficiency wind mass-loss rate parameter is $\eta=0.12$. Note the different scales of the three panels.}
\label{fig:HD_38529c_TEST5}
\end{figure}

To further reveal the dependence of the fate of the planet on the properties of its orbit we examine the role of eccentricity. We take the planet HD~38529~c with an observed eccentricity of $e_0=0.36$ and search for the initial eccentricity, $e_{\rm n,0}$, that would allow the star to engulf the planet during the AGB phase. We make the calculations for one value of the wind mass-loss rate parameter $\eta=0.12$, and find that, keeping all other observed parameters unchanged, an initial eccentricity of $e_{\rm n,0} \ga 0.6$ would have allowed a CEE to take place. We present the results in Table \ref{tab:eccentricity_change}, where the meanings of the different variables are as in Table \ref{tab:FiveExoplanets}. In Fig. \ref{fig:HD_38529c_TEST5} we present the evolution of stellar radius, semi-major axis, periastron distance, and eccentricity, in the post-main sequence phases. 
\begin{table*}
\centering
\begin{tabular}{|c|c|c|c|c|c|c|}
\hline
\multicolumn{4}{|c|}{HD 38529 c}  & \multicolumn{3}{c|}{Outcome: No CEE or $M_{\rm core};M_{\rm env}$ } \\ \hline
 $M_{\ast, 0}$ & $M_{\rm p}$ & $a_0$ & $e_0$ & $e_{\rm n,0}=0.4$ & $e_{\rm n,0}=0.5$ & $e_{\rm n,0}=0.6$ \\ \hline
 1.48 & 23.7 & 793 & 0.36 & No CEE & No CEE & 0.563;0.715 \\ \hline
\end{tabular}
\caption{Examining for the eccentricity of the orbit of HD~38529~c that would bring it to form a CEE during the AGB phase of its parent star. The first four columns in the second row are the observed values where units are as in Table \ref{tab:FiveExoplanets}. The last three columns indicate the outcomes had the eccentricity of the orbit been larger, keeping all other observed properties unchanged. In all simulations the wind mass-loss rate parameter is $\eta=0.12$. For $e_{\rm n, 0}=0.6$ the planet enters a CEE, and we list the core and envelope masses (in $M_\odot$) at the onset of the CEE. }
\label{tab:eccentricity_change}
\end{table*}

The result of this simple study is expected, namely, a higher eccentricity for a given semi-major axis, which gives a smaller periastron distance, increases the likelihood of engulfment. However, it is not a straightforward evolution, because as we see in Fig. \ref{fig:HD_38529c_TEST5} the eccentricity and semi-major axis of the orbit decrease already during the upper RGB phase of the parent star (see also Fig. \ref{fig:beta_Picc_TEST3} for the planet beta~Pic~c). 
The periastron distance $a_p=(1-e)a$, though, increases. 
As the initial eccentricity $e_{\rm n, 0}$ increases, the decrease in the semi-major axis and eccentricity on the RGB becomes more significant.
The evolution with $e_{\rm n, 0}=0.6$ has a smaller  semi-major axis than the other two cases when the system leaves the RGB. This smaller semi-major axis makes tidal interaction on the AGB stronger, and the system is more likely to enter a CEE. 

\begin{figure}
\includegraphics[trim= 5.0cm 5.5cm 2.5cm 6.5cm,clip=true,width=0.6\textwidth]{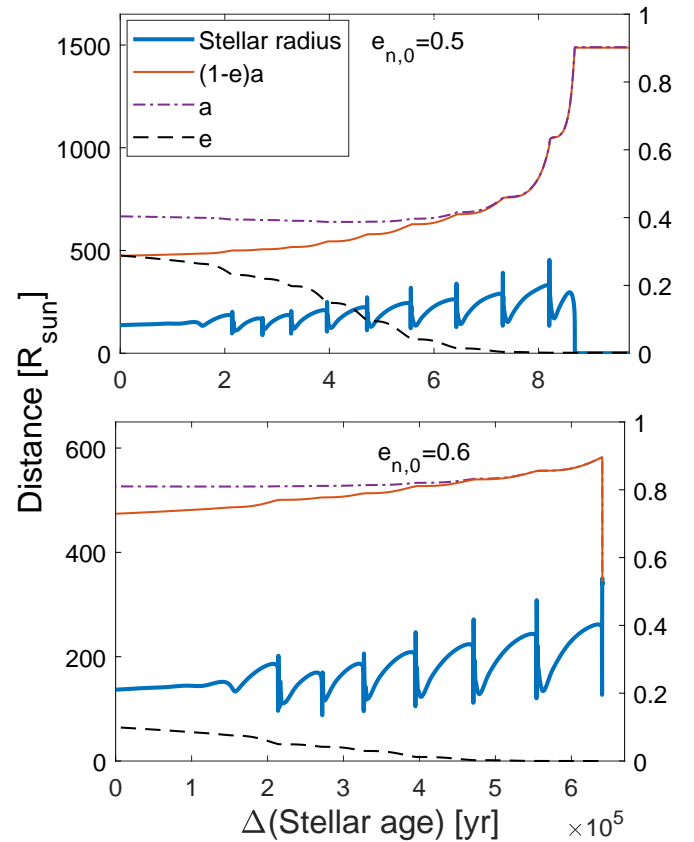}
\caption{Zooming on the final AGB evolution of the two lower panels of Fig. \ref{fig:HD_38529c_TEST5}. {{{{ The spikes in the blue line (stellar radius) are helium-shell flashes. }}}} The two panels start and $t=3.06 \times 10^9 \yr$. Note the different scales of the two panels. }
\label{fig:ZoomedToHD38529c}
\end{figure}

In Fig. \ref{fig:ZoomedToHD38529c} we zoom on the last million years or so. As in the evolution of  beta~Pic~c (Fig \ref{fig:ZoomedToAGB}), engulfment occurs following a stellar expansion as a result of helium-shell flash, when the orbit is already circular. 
     
\section{Summary}
\label{sec:Summary}

The main goal of this study is to better understand the engulfment of planets during the RGB and AGB phases of their parent stars, in particular in relation to the possibility that planets shape the outflow of some AGB progenitors of elliptical PNe. 
{{{{ Planets (and BDs) can affect the mass-loss geometry of their parent AGB (or RGB) star to form an elliptical PN by spinning-up the envelope and/or by exciting waves in the envelope (section \ref{sec:intro}). The spinning-up process takes place mainly as the planet tidally interacts with the envelope and when it enters the envelope, while excitation of waves takes place mainly when the planet is deep inside the envelope. Both of these processes influence the mass-loss geometry mainly by affecting the formation of dust on the surface (section \ref{sec:intro}). }}}}

Our approach here followed earlier studies (e.g., \citealt{NordhausSpiegel2013, SabachSoker2018a}) in following the evolution of confirmed exoplanets (and BDs). We specifically focused on planets that have orbits with semi-major axis in the range of $1 \la a_0 \la 20 \AU$ and high eccentricities. We examined five systems,  from the Extrasolar Planets Encyclopaedia; (exoplanet.eu; \citealt{Schneideretal2011}) and from  \cite{Maireetal2020} that fit our requirements. To study their evolution we used the stellar evolutionary code   \textsc{MESA-binary}.

We also followed \cite{SabachSoker2018a} and assumed that low mass stars that do not acquire angular momentum from a companion (Jsolated stars) have a much lower wind mass-loss rate during their RGB and AGB phases than the commonly used value ($\eta =0.5$ in equation \ref{eq:Reimers}). We summarised the fate of the planets in Table \ref{tab:FiveExoplanets}. 

We found that out of the five systems, one system, HIP~75458, enters a CEE during the RGB phase of the parent star for all values of $\eta$ (Figs. \ref{fig:HIP_75458b_TEST3} and \ref{fig:HIP_75458b_Zoom}). The planet removes the envelope and leaves a bare helium core that will evolve to form a helium white dwarf. 

Only in one system the planet, beta~Pic~c, enters the envelope of its parent star during the AGB phase. For that to occur, we had to reduce the wind mass-loss rate by a factor of about 4 ($\eta \la 0.12$; table \ref{tab:FiveExoplanets}). 
The four other systems do not enter a CEE phase even for the lowest value of $\eta$. 

Overall, our study of eccentric planetary systems strengthens the early conclusion of \cite{SabachSoker2018a} that was based on circular orbits and used a simple tidal interaction formula. The conclusion is that to have a non-negligible fraction of AGB stars that engulf planets we should consider a lower wind mass-loss rates of Jsolated stars. 

We also made a test on the influence of the eccentricity. Keeping all other parameters at their observed value, we examined for what eccentricity of its orbit the planet HD~38529~c would enter a CEE with its parent star during the AGB phase. The observed value of the eccentricity is $e_0=0.36$. We found that we need to increase the eccentricity to a value of $e_{\rm n,0} \ga 0.6$ for AGB engulfment to take place (table \ref{tab:eccentricity_change}).   

In the cases where we do have engulfment on the AGB, the evolution involves some decrease in eccentricity and in the semi-major axis on the upper RGB phase, although the periastron distance $(1-e)a$ increases (Figs. \ref{fig:beta_Picc_TEST3} and \ref{fig:HD_38529c_TEST5}). 
The final AGB engulfment takes place after a large envelope expansion as a result of a helium-shell flash (Figs. \ref{fig:ZoomedToAGB} and \ref{fig:ZoomedToHD38529c}). 

The next step is to conduct a thorough statistical study. 
However, the number of relevant confirmed exoplanets with semi-major axis of $1 \AU \la a_0 \la 20 \AU$ around potential progenitors of PNe (stars with initial masses of $M_{\ast, 0} \ga 1M_\odot$) is too low to conduct a meaningful statistical study. The uncertainty in the wind mass-loss rate on the RGB, and in particular on the AGB, adds to the uncertainty of such a study. Nonetheless, we encourage future studies to follow the evolution of exoplanets as they are discovered, to better learn about their degree of significance in influencing the post-RGB evolution and/or in potentially shaping elliptical PNe.
         
\acknowledgments

{{{{ We thank an anonymous referee for several corrections and for improving our manuscript. We thank Efrat Sabach for valuable comments. }}}}
This research was supported by a grant from the Israel Science Foundation. We completed this work while the Technion was closed due to the Coronavirus (COVID-19).

\pagebreak

\end{document}